\documentclass[conference]{IEEEtran}

\usepackage{xcolor,soul,framed} %,caption

\colorlet{shadecolor}{yellow}
\usepackage[pdftex]{graphicx}
\graphicspath{{../pdf/}{../jpeg/}}
\DeclareGraphicsExtensions{.pdf,.jpeg,.png}

\usepackage[cmex10]{amsmath}
\usepackage{mathabx}
\usepackage{array}
\usepackage{mdwmath}
\usepackage{mdwtab}
\usepackage{eqparbox}
\usepackage{url}
\usepackage{cite}
\usepackage{multirow}
\usepackage{makecell}
\usepackage{subcaption}
\usepackage{algorithm}
\usepackage{algpseudocode}
\usepackage{pgffor}
\usepackage{tabularx}  % For flexible column width
\usepackage{booktabs} % For better table formatting

\begin{document}
%\bstctlcite{IEEEexample:BSTcontrol}
%\title{Consensus Clustering for the Identification of Coherent Regions in IBR Dominant Grid\\
 \title{Consensus Clustering for the Identification of Coherent Regions with Varied Generation Mix\\
}
    %by Clusters of Composite Load Models}
\author{\IEEEauthorblockN{Kiran~Kumar~Challa,}
\IEEEauthorblockA{\textit{Dept. of Electrical \& Computer Engg.} \\
\textit{Iowa State University}\\
Ames, IA, USA \\
kiranc@iastate.edu}
\and
\IEEEauthorblockN{Alok~Kumar~Bharati,~\IEEEmembership{Senior Member,~IEEE,}}
\IEEEauthorblockA{Richland, WA, USA \\
\textit{\textit{alok.bharati@outlook.com}   }\\
 \\
}
\and
\IEEEauthorblockN{Venkataramana Ajjarapu,~\IEEEmembership{Fellow,~IEEE}}
\IEEEauthorblockA{\textit{Dept. of Electrical \& Computer Engg.} \\
\textit{Iowa State University}\\
Ames, IA, USA \\
vajjarap@iastate.edu}

}

% ====================================================================
\maketitle

% === ABSTRACT ====================================================================
% =================================================================================
\begin{abstract}
%\boldmath
With a steady increase in the inverter technology integration to the grid, frequency response of the large interconnection system becomes more unpredictable. This leads to a significant change in the boundaries of the coherent region, which highly depends on the changing disturbance locations and operating conditions. While most of the existing coherency identification is based on a single large generator outage, it is important to identify these boundaries in view of wide range of disturbances. With large amount of inverters in the system, there is increase in the dynamic interactions of the various grid components leading to a need for such boundary identification. This paper presents the multi-view consensus algorithm to identify coherency in the case of variable grid operating conditions and wide range of disturbances. The proposed approach is demonstrated by identifying the coherent regions in the miniWECC 240 bus test system.

\end{abstract}

% === KEYWORDS ====================================================================
% =================================================================================

% For peer review papers, you can put extra information on the cover
% page as needed:
% \ifCLASSOPTIONpeerreview
% \begin{center} \bfseries EDICS Category: 3-BBND \end{center}
% \fi
%
% For peerreview papers, this IEEEtran command inserts a page break and
% creates the second title. It will be ignored for other modes.
\IEEEpeerreviewmaketitle

% ====================================================================
% ====================================================================
% ====================================================================

% === I. INTRODUCTION =============================================================
% =================================================================================
\section{Introduction}
Coherency identification is essential in modern grid operations to ensure that generation meets the demand, frequency remains stable, and tie-line flows are controlled, preventing major disturbances and enabling seamless grid operation \cite{9451565}. With the increasing inverter based resource (IBR) integration, decreasing system inertia makes it critical to maintain inertial support in every coherent region. Coherency identification also has several applications such as dynamic model reduction \cite{9141428,7866837}, control islanding \cite{8708689,9091154}, and wide area control \cite{osti_1094827,8937835}.

Several model-based and measurement-based approaches are presented in literature to identify coherent generators in a large interconnected system. Model-based approaches rely on the linear/nonlinear simulation of the power system and uses the time domain responses for the CI. The concepts of slow coherency \cite{SlowCoh_CI}, electromechanical distance in the transient period \cite{EleMechDist_CI} and relative coupling between the generators \cite{Couple_CI} are presented using the linearized power system model. These methods suffer accuracy due to the model approximations.
These methods are associated with the inter-area oscillations which suffer inefficient when there are significant inter-area oscillations \cite{RT_CI}. Measurement-based approaches employ the PMUs to identify coherency through Fourier transform \cite{FT_PMU} and dynamic model decomposition \cite{DMD_PMU} of the frequency responses. These approaches are unreliable in the changing grid conditions due to their sensitivity to noises, spurious signal components and depends on the type of disturbance. The online measurement based methods do have limitations on the data-collection infrastructure, excessive computational burden, and bandwidth requirements that affect coherency grouping's success. While all the existing coherency identification approaches consider the occurrence of a single large contingency during the process, the clustering results under different contingencies such as disturbance location and type of disturbance have significant impact on the coherency boundaries.

With increasing penetration of IBRs, frequency deviation in large-interconnected power systems is unpredictable. Coherency between the bus frequencies also changes with generation mix in the system at the time of disturbance and network which makes the identification of coherent regions more difficult. We propose to identify the coherent regions for a set of contingencies covering the location of generators, size of generators, type of source for a given operating condition. We use multi-view consensus clustering that aims to integrate information from different frequency-related contingencies to arrive at a single, unified clustering solution. We propose to utilize the advantage of effective ensemble learning presented in \cite{EnCluster} to achieve the unified clustering. Ensemble learning-based clustering is a technique where multiple clustering results are combined into a single, more reliable, and stable clustering solution. The goal is to find a consensus among different views that enhances the quality of the clustering results. This approach can be particularly effective when each contingency provides complementary information about the underlying coherency between the bus frequencies. 

\section{Variability in Coherency}
\begin{figure}[!h]
\centering
  \includegraphics[width=0.75\linewidth]{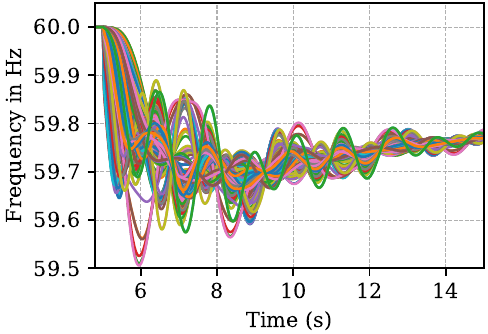}
  \caption{Frequency responses for a trip of generator at bus-4131}\label{Hydro4131}  
\end{figure}

\begin{figure}[!h]
\centering
  \includegraphics[width=0.75\linewidth]{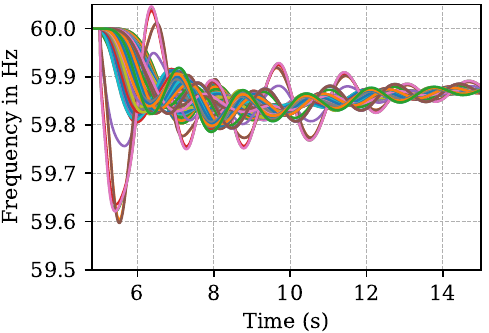}
  \caption{Frequency responses for a trip of generator  at bus-5032}\label{Coal5032}  
\end{figure}

%Frequency response in the case of miniWECC 240 system is studied for generator outage contingencies at different locations and different capacities. The miniWECC 240-bus system is a reduced-order power system model representing a simplified version of the Western Electricity Coordinating Council (WECC) grid.  In this model, the total generation capacity is approximately 291 gigawatts (GW), with about 59 GW (roughly 20\%) attributed to grid-following inverter-based resources (IBRs), including utility-scale photovoltaic (PV) resources, wind, and distributed PV. Frequency response of this system is observed for two different generator outages to understand the variation in the coherency with the change in the disturbance. Generator outage disturbance are created with single large generator in the system at bus 4131 of 7418 MW capacity and another large generator at bus 5032 of 3646 MW capacity. Fig. \ref{Hydro4131} and Fig. \ref{Coal5032} present the frequency with a 7418 MW (at bus 4131) hydro generator and 3646MW (at bus 5032) coal generator unit. 
The penetration of varied generation resulted in the multiple small generations being interconnected to the grid instead of larger generators in the case of synchronous generator-based resources. These generators do spread across the large interconnected system and impact the coherency of the grid. Identification of coherency using the system dynamics during the largest generator outage becomes infeasible as the new interconnected generators are not larger in size and largest synchronous generators are retiring. Importance of the online coherency identification is emphasized in \cite{WECC_FC}. It is also identified in \cite{WECC_FC} that the coherency changes with each disturbance in the system. Though the online coherency identification helps in the control, off-line identification in view of multiple disturbances is critical in planning of the real power reserves. Frequency responses in the case of different generator outages is studied using miniWECC 240 system to demonstrate the critical aspects of coherency variation and results are presented in Fig. \ref{Hydro4131}
 and Fig. \ref{Coal5032}. The miniWECC 240-bus system is a reduced-order power system model representing a simplified version of the Western Electricity Coordinating Council (WECC) grid.  In this model, the total generation capacity is approximately 291 gigawatts (GW), with about 59 GW (roughly 20\%) attributed to grid-following inverter technology, that includes variable output generation. Frequency response of this system is observed for two different generator outages to understand the variation in the coherency with the change in the disturbance. Generator outage disturbances are created with a single large generator in the system at bus 4131 of 7418 MW capacity and another large generator at bus 5032 of 3646 MW capacity. Fig. \ref{Hydro4131} and Fig. \ref{Coal5032} present the frequency with a 7418 MW (at bus 4131) hydro generator and 3646MW (at bus 5032) coal generator unit.

It is evident from these responses that the frequency nadir, ROCOF and time of frequency nadir are completely different. The trip of generator at 5032 causes a similar frequency nadir in spite of the amount of generator outage is almost half. Similarity index matrix for the first disturbance is compared with that of the second disturbance, and it is observed that the coherency between the bus frequencies is different in the case of the two disturbances. This also indicates that the frequency is nonlinear and unpredictable in a system with IBR penetrations which leads to the variations in the coherency. This similarity matrix for the wide range of generator outages will be utilized in the clustering algorithm to identify the coherent regions. We propose to use multi-view consensus clustering algorithm to identify the coherent regions in view of varying coherency for different generator outages.

% ====== 3D K-mean Clustering Algorithm
\section{Consensus Clustering for Coherent Region Identification} \label{ConsAlgo}

\begin{figure*}[!h]
\centering
  \includegraphics[width=0.85\linewidth]{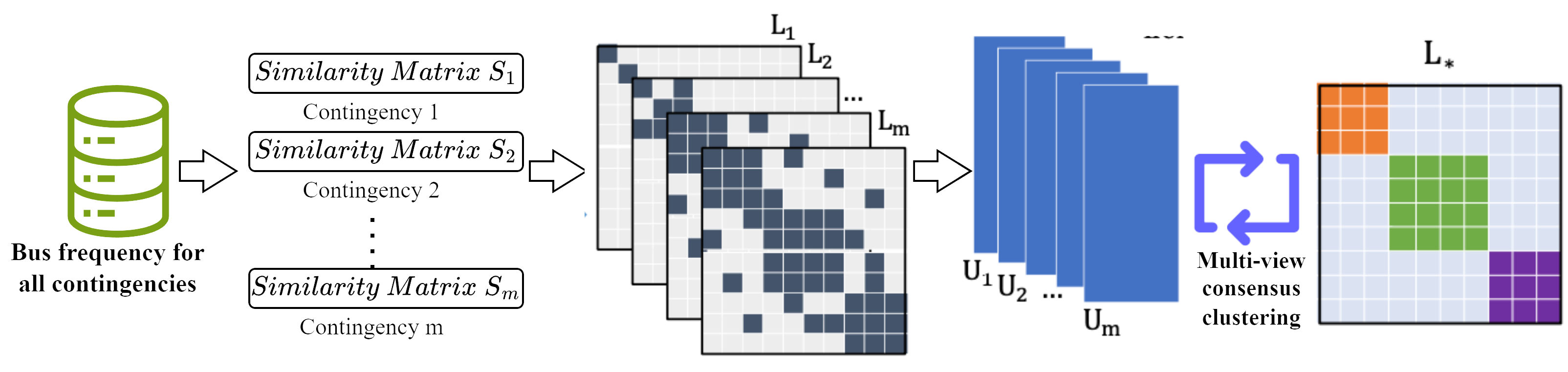}
  \caption{Illustration of consensus clustering algorithm}\label{FC_Algo_CC}  
\end{figure*}

Identifying frequency coherent regions that correspond to multiple generator outages at different locations of a large interconnected network required the clustering of the nodes in view of multiple contingencies. Spectral clustering is a powerful graph-based clustering technique that is widely used to extract the eigenvalues of a node similarity matrix to get the low dimensional information before applying a standard clustering method. We propose to use consensus learning method to the spectral clustering approach to identify the unified clusters in view of multiple disturbances instead of using the clustering results of each disturbance. The goal is to find a consensus among different views that enhances the quality of the clustering results through the process outlined in Fig. \ref{FC_Algo_CC}. This consensus approach is generic and applicable for any multi-view aspects such as variability in operating conditions, different measurements considered for coherency identification, variability in generation mix, and different base clustering approaches. This approach learn the robust representation of graph Laplacian in each base spectral clustering of the disturbance scenario. Multiple generator outage contingencies (\textit{m} number of contingencies) are created in the selected test system, and the frequency response at all the buses (\textit{$N_b$} number of buses) for each contingency is recorded in the data repository. This data repository is used to identify the similarity matrix of every contingency in stage 1 as shown in Fig. \ref{FC_Algo_CC}. In the next stage, coherency for each disturbance is identified using the similarity matrix. Final coherent regions are identified using the multi-view consensus clustering of the individual disturbance results. Step by step procedure for the clustering using the time-varying responses of the frequencies for the selected contingencies using the consensus clustering algorithm adopted is presented below,
%In a synchronous generator dominant power system, coherent areas are determined by the frequency boundaries determined after a single large power imbalance [IITDreference]. With increasing penetration of IBRs, frequency deviation in a large-interconnected power systems is unpredictable and depends on various factors such as, the location of the disturbance, type of the generator outage, generation mix in the system at the time of disturbance and network. Coherency between the bus frequencies also changes with the location of disturbance which makes the identification of coherent regions to be more difficult. We propose to identify the coherent regions for a set of contingencies covering the location of generators, size of generators, type of source for each representative operating condition defined in sections 2.1. We use multi-view consensus clustering that aims to integrate information from different frequency related contingencies to arrive at a single, unified clustering solution. The goal is to find a consensus among different views that enhances the quality of the clustering results. This approach can be particularly effective when each contingency provides complementary information about the underlying coherency between the bus frequencies. Following are the steps involved in the consensus clustering and the graphical view of the algorithm is presented in Figure 7, 
\begin{itemize}
    \item[1]  \textbf{Calculation of similarity matrix for selected contingencies}: 
    Similarity matrix (S) is essential for every contingency for the calculation of graph Laplacian matrix. We calculate \textit{S} matrix in this step, where \textit{S} is of the order $N_b \times N_b$ and relates the coherency between the bus frequencies through the Pearson correlation coefficient. The Pearson correlation coefficient ($r_{ij}$ in S) is a statistical measure that quantifies how much the two frequencies at different buses ($f_i$ and $f_j$) are related to each other.
    \begin{align}
        r_{ij}=\frac{\sum_{k=1}^{n}(f_{ik}-\overline{f_l})(f_{jk}-\overline{f_j})}{\sqrt{\sum_{k=1}^{n}(f_{ik}-\overline{f_l})^2}\sqrt{\sum_{k=1}^{n}(f_{jk}-\overline{f_j})^2}}
    \end{align}
    Where  $f_{ik}$ and $f_{jk}$ are the $k^{th}$ values of the frequencies $f_i$ and $f_j$ respectively.  $\overline{f_i}$ and $\overline{f_j}$ are the mean of all values of $f_i$ and $f_j$. \textit{n} is the total number of data points in $f_i$ or $f_j$.  The value of ‘$r_{ij}$’ ranges between -1 and 1, where $r_{ij}=1$ means a perfect positive linear correlation, $r_{ij}=-1$ means a perfect negative linear correlation, and $r_{ij}=0$ means no linear correlation. For a set of ‘m’ contingency data, calculate the similarity matrices $S=S_1,S_2,….,S_m$.

    \item[2] \textbf{ Calculate the Laplacian matrix}: Severity of the frequency variations and similarity between the bus frequencies are mathematically represented in a Laplacian matrix for every $S$ matrix. The degree matrix (D) which indicates the severity of the frequency variation is a diagonal matrix with entries $d_1,d_2,…….,d_j,….d_{b}$, as defined in \ref{dMat}, along its diagonal.
    \begin{align}
        d_j = \sum_{k,k\neq j}^{N_b} S(j,k),\ \ \ S \in R^{N_b \times N_b} \label{dMat}
    \end{align}
    Calculate the graph Laplacian matrices using $L=D-A$ from the Similarity matrix (S) and degree matrix (D) for all the generator outages considered in the analysis.

    \item[2] \textbf{Calculation of low dimensional representation of Laplacian}: Laplacian matrices $\mathcal{L}=L_1,L_2,….,L_m$ corresponding to S are calculated using the spectral clustering algorithm presented for the voltage contingency clustering algorithm presented above. Low dimensional vector representation $\mathcal{U}=U_1,U_2,….,U_m$ of the original Laplacian matrices $\mathcal{L}$ is obtained through the following formulation.
    \begin{align}
        \max_{U} tr(U^T L U),\ \ \ \ s.t\ \ \ U^T U=I
    \end{align}
    Since $\mathcal{U}$ is calculated from $\mathcal{L}$, $\mathcal{U}$ contains the intrinsic data structures explored by $\mathcal{L}$. 
    \item[3] \textbf{Calculation of consensus Laplacian matrix and final clusters}: $U_1,U_2,….,U_m$ can be utilized to calculate the initial consensus Laplacian ($L_*$) by setting $L_*=\sum _{i=1}^{m} U^T_i U_i)$, which can enhance the block-diagonal structure of the learned representative graph Laplacian and better uncover the cluster structure. By incorporating the representative graph Laplacian learning and the optimization of each base spectral clustering algorithm into a joint framework for consensus clustering, the proposed multi-view consensus clustering method is formulated as follows,
    \begin{equation}
        \max_{U_1,U_2,….,U_m,U_*} \sum_{i=1}^m tr(U^T_i L_i U_i)+\alpha\ tr(U^T_* L_* U_*)
    \end{equation}
     \begin{align}
    s.t \ \ U^T_i U_i=I;  & \ \   U^T_* U_*=I,\nonumber \\
        \ \ L_*=\sum _{i=1}^{m} U^T_i U_i);\ \  
        \forall 1\leq i \leq m \nonumber
    \end{align}
    Where $\alpha>0$ is a balance parameter. $\alpha$ is the learning rate which essentially controls how fast the base spectral is optimized. We select an adaptive setting proposed in [11] to choose the parameter automatically in each iteration to optimize the objective function.
    \item[4] The above optimization problem is not convex when $U_1,U_2,….,U_m$ and $U_*$ are optimized simultaneously. To optimize the objective function, the above optimization problem is solved in 2 states: updating $U_*$ and $L_*$ while fixing $U_1,U_2,….,U_m$, and updating $U_1,U_2,….,U_m$ while fixing $U_*$ and $L_*$.
\end{itemize}
Instead of directly using the clustering results that may exist mis-clustering caused by single disturbance, the proposed consensus clustering algorithm uses the graph Laplacian of spectral clustering to generate the base clusterings, then optimize each base spectral clustering unitizing the consensus Laplacian matrix until convergence, finally finds an optimal cluster structure in a low-dimension based on the consensus graph Laplacian. The optimization process is also accelerated using the adaptive parameter settings to achieve better performance and fast convergence.

% ====== End og 3D K-mean Clustering Algorithm

% === II. Harmonically-Terminated Power Rectifier Analysis ========================
% ==================================================================

\section {Coherency Identification Results}
The coherency analysis is performed on the MiniWECC 240-bus transmission network, which represents a reduced but dynamically representative model of the Western Interconnection. This system includes a diverse set of generators, transmission corridors, and load pockets, making it well suited for evaluating disturbance propagation and coherency behavior across a large-scale grid. Its realistic topology and modal characteristics allow meaningful interpretation of coherency patterns under various disturbance scenarios.

Initially, coherency identification is carried out using the frequency responses obtained from a single large disturbance created at Bus 4131, and spectral clustering is applied to group the buses based on their dynamic similarity. The number of buses assigned to each of the 10 coherent regions resulting from this single-disturbance case is summarized in Table.\ref{Coherency1Dist}. Notably, the first coherent region contains approximately 35\% of the system—91 buses—indicating that a substantial portion of the network exhibits nearly identical frequency behavior under this disturbance. To enhance the robustness of the coherency assessment, multiple generator outage scenarios are additionally considered, as listed in Column 1 of Table.\ref{Coherency1Dist}. For each outage case, the corresponding number of buses in all ten coherent regions is reported in Columns 2–11. The comparative results clearly show that the single disturbance at Bus 4131 does not propagate meaningful frequency variations throughout the entire network. Consequently, many buses located electrically or geographically far from the disturbance exhibit very small frequency deviations and are therefore grouped into the same coherent region, even though their underlying dynamic behavior may differ during other credible contingencies.

This observation highlights a key limitation of relying on a single disturbance for coherency identification: it fails to sufficiently excite the full system and cannot reveal distinct dynamic signatures across all regions. The results reinforce the necessity of using multiple large generator outage events—or a diverse set of credible contingencies—to accurately uncover true coherent areas in the network and avoid mis-classification driven by insufficient disturbance diversity.

\begin{table}[h]
    \centering
    \caption{Coherency with single large disturbance}
    \begin{tabular}{*{11}{c}} % Creates 10 columns with center alignment
        \toprule
       Region No &1 & 2 & 3 & 4 & 5 & 6 & 7 & 8 & 9 & 10 \\
       \midrule
       Gen at 1431 &91 & 64 & 13 & 9 & 26 & 5 & 6 & 5 & 6& 18 \\  
       \midrule
       Gen at 1232 &101 & 43 & 15 & 15 & 11 & 35 & 3 & 14 & 2& 4 \\ 
       \midrule
       Gen at 2523 &73 & 52 & 73 & 16 & 15 & 3 & 3 & 3 & 3& 2 \\ 
       \midrule
       Gen at 5031 &11 & 60 & 39 & 26 & 18 & 36 & 21 & 22 & 5& 5 \\ 
        \midrule
      Consensus Cl & 41 & 49 & 6 & 19 & 34 & 27 & 5 & 11 & 27& 24 \\  
        \bottomrule
    \end{tabular}
    \label{Coherency1Dist}
\end{table}

\begin{figure}[!h]
    \centering
    \begin{subfigure}{0.49\linewidth}
        \centering
        \includegraphics[width=\linewidth]{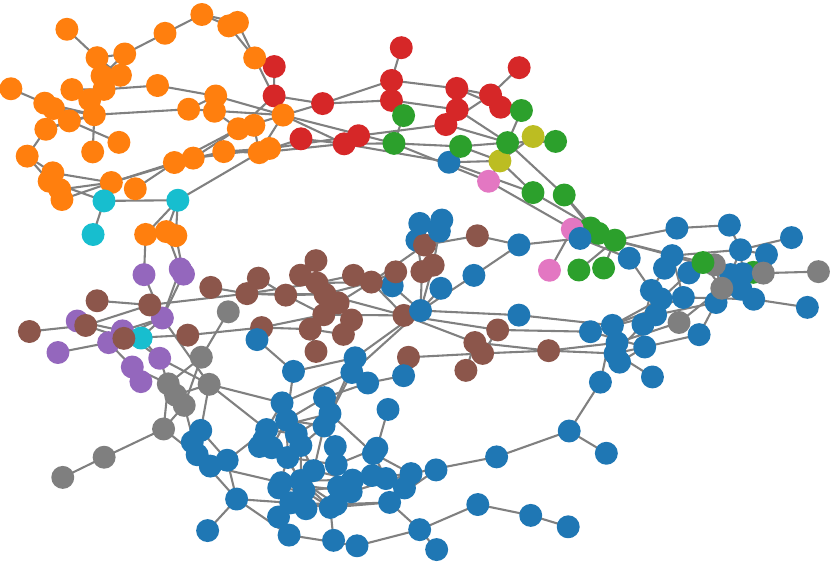}
        \caption{Gen at 1232}
        \label{dist1}
    \end{subfigure}
    \begin{subfigure}{0.49\linewidth}
        \centering
        \includegraphics[width=\linewidth]{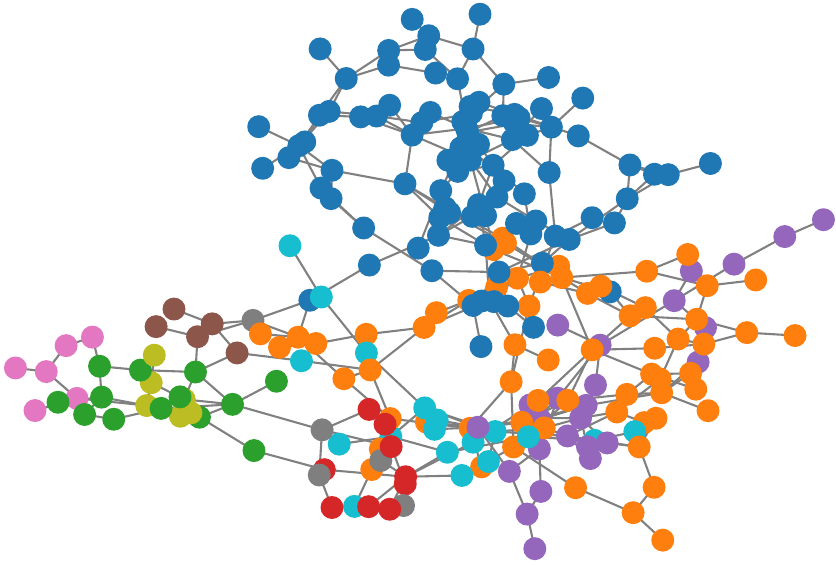}
        \caption{Gen at 4131}
        \label{dist2}
    \end{subfigure}
    \begin{subfigure}{0.49\linewidth}
        \centering
        \includegraphics[width=\linewidth]{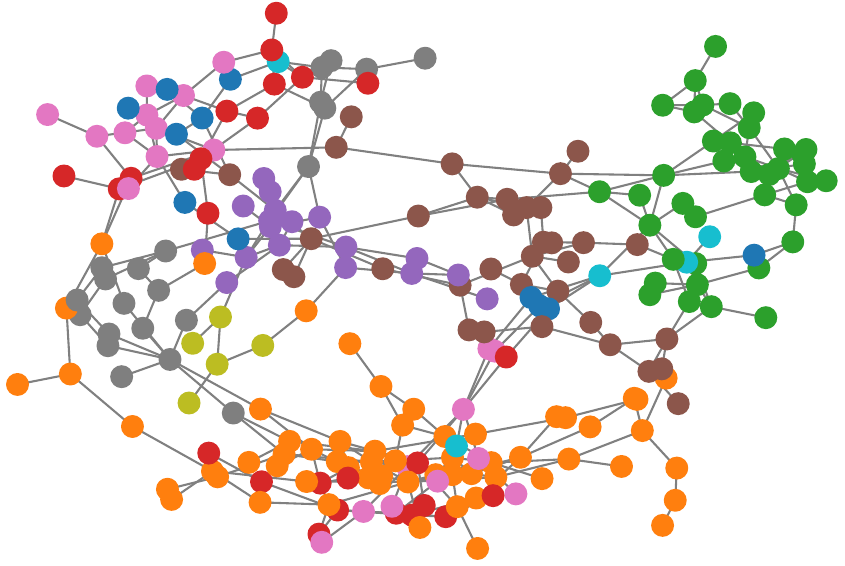}
        \caption{Gen at 5031}
        \label{dist3}
    \end{subfigure}
    \begin{subfigure}{0.49\linewidth}
        \centering
        \includegraphics[width=\linewidth]{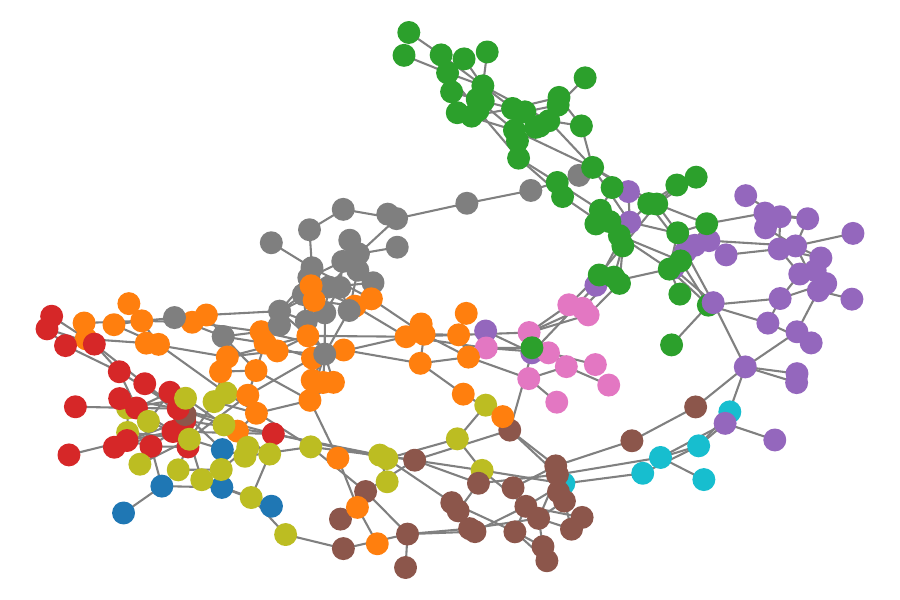}
        \caption{Consensus Results}
        \label{dist_all}
    \end{subfigure}
    \caption{Coherent regions for different generator outages and proposed method}
    \label{CoherentResponses}
\end{figure}

To identify coherent regions that remain valid across a broad spectrum of operating conditions and disturbances, ten distinct large-generator outage scenarios were created within the MiniWECC test system. Two high-capacity generators were selected from each major geographical area to ensure that the imposed contingencies were sufficiently severe and spatially diverse, thereby inducing measurable frequency separations reflective of true dynamic coherency. For every disturbance, the bus-level frequency responses were recorded, forming a comprehensive multi-contingency frequency response dataset.

This frequency response repository serves as the input to the consensus-based coherency identification framework described in Section \ref{ConsAlgo}. For each contingency, the frequency trajectory at every bus is mapped into a similarity matrix, from which the corresponding graph Laplacian is constructed. These Laplacian matrices collectively represent multiple “views” of the system’s dynamic behavior. The proposed consensus optimization then integrates these views to extract a unified clustering solution that is simultaneously valid for all disturbances, thereby overcoming the limitations of single-contingency coherency assessment.

\begin{figure}[h]
    \centering
    \begin{subfigure}{0.49\linewidth}
        \centering
        \includegraphics[width=\linewidth]{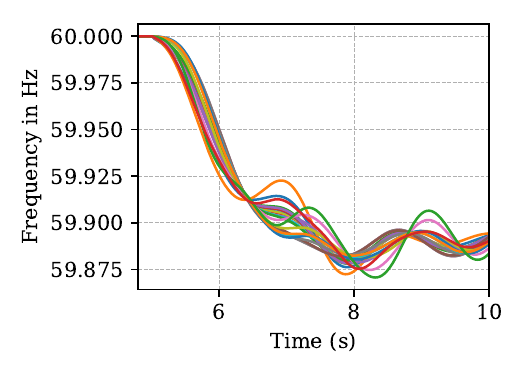}
        \caption{Gen at 1232}
        \label{gen1232}
    \end{subfigure}
    \begin{subfigure}{0.49\linewidth}
        \centering
        \includegraphics[width=\linewidth]{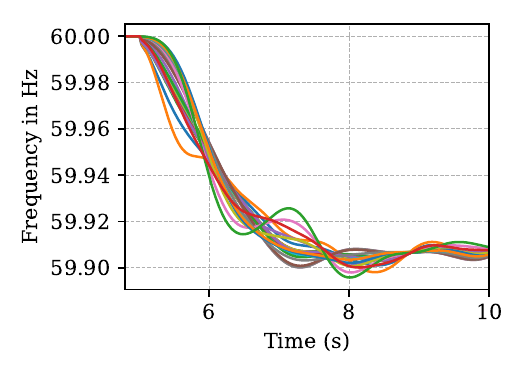}
        \caption{Gen at 2630}
        \label{gen2630}
    \end{subfigure}
    \begin{subfigure}{0.49\linewidth}
        \centering
        \includegraphics[width=\linewidth]{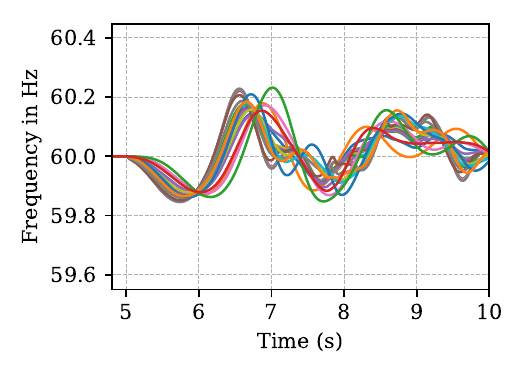}
        \caption{Gen at 5031}
        \label{gen5031}
    \end{subfigure}
    \begin{subfigure}{0.49\linewidth}
        \centering
        \includegraphics[width=\linewidth]{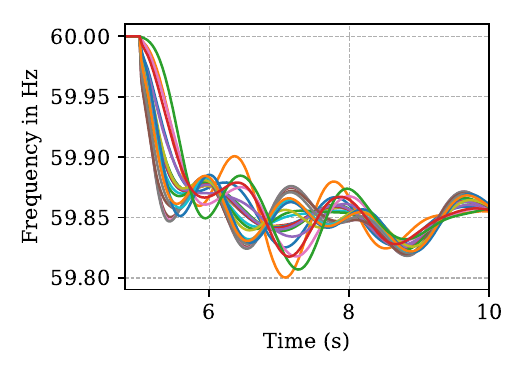}
        \caption{Gen at 4132}
        \label{gen4132}
    \end{subfigure}
    \caption{Frequency response of a coherent region for different generator outages}
    \label{CoherencyResults}
\end{figure}

The number of coherent groups is determined using the Silhouette Score (SS), computed for a range of possible cluster counts. The optimal number of coherent regions is selected as the value maximizing the SS, which for the MiniWECC system is found to be ten. The coherency identification outcomes using the proposed consensus algorithm are illustrated in Fig. \ref{CoherencyResults}. For comparison, Figs. \ref{dist1}–\ref{dist3} depict the single-view clustering results under three representative generator outages. These individual disturbances fail to induce significant frequency divergence in electrically distant portions of the network, leading to the formation of an oversized coherent region containing more than 100 buses—an artifact of insufficient system excitation.
In contrast, the proposed consensus algorithm synthesizes information from all ten outages, enabling the extraction of coherent areas that truly persist across multiple credible disturbances. The final cluster memberships, including the number of buses in each coherent region, are presented in the last row of Table \ref{Coherency1Dist}. Unlike the single-contingency (single-view) approaches—which produce one dominant coherent region and several small regions with only a few buses—the consensus-based method yields a much more balanced and physically meaningful partitioning of the network into multiple coherent regions.

%Number of buses in every coherent region is mentioned in the brackets of the first column along with the region number. Details of the bus numbers in the coherent region are provided in the second column of the table. Unlike the single view clustering presented in the previous case where 204 buses are part of a single coherent region and three regions with a single bus only, the proposed identification process could evenly distribute the network into multiple coherent regions.

To further validate the coherency identification results, frequency trajectories corresponding to a specific coherent region are plotted for multiple contingencies in Fig. \ref{CoherentResponses}. The figure shows the frequency responses of Region 9 under four representative generator outages at buses 1232, 2630, 5031, and 4132. As illustrated, the buses within this coherent region exhibit highly similar dynamic profiles across all contingencies. In particular, the overall shape of the frequency trajectories and the rate-of-change of frequency (RoCoF) remain closely aligned, confirming that these buses respond in a coordinated manner despite the disturbances originating from different locations.

However, unlike systems dominated by synchronous generation—where the frequency within a coherent region typically follows an almost identical and smooth pattern—the MiniWECC system contains a substantial share of inverter-based resources (IBRs). The presence of fast-acting inverter controls introduces noticeable variations and higher oscillatory components in some bus-level responses within each coherent region. These localized oscillations stem from the nonlinear and rapid control actions of grid-forming or grid-following inverters, which tend to react more quickly than synchronous machines and thus introduce mild divergence from perfectly matched trajectories. Nevertheless, these deviations remain within the bounds of coherent behavior when evaluated through the similarity-based clustering framework.

The proposed consensus clustering approach effectively captures these nuances and still reliably identifies the coherent regions in the presence of mixed resource types, including systems with high IBR penetration. This demonstrates the robustness of the method and its capability to generalize across heterogeneous dynamic behaviors. Consequently, the identified coherent regions can be leveraged to determine region-specific requirements such as real-power reserve allocation, synthetic inertia contributions, and fast frequency response needs—particularly valuable under changing operating conditions and evolving resource mixes.

\section{Conclusion}
%This paper focuses on the identification of coherent regions in a high IBR penetrated grid in view of multiple disturbances. Coherent regions are identified in a large interconnected system by considering the frequency responses during multiple generator outages at diverse locations in the system with multiple types generation types. Consensus clustering algorithm is presented to identify a single set of coherent regions valid for all the disturbances considered. Proposed algorithm is implemented to identify coherency in miniWECC 240 bus test system having 20\% of its generation from inverter-based resources.

This paper focuses on the identification of coherent regions in future grid with varied generation mix in view of multiple disturbances. Coherent regions are identified in a large interconnected system by considering the frequency responses during multiple generator outages at diverse locations in the system with multiple types generation types. Consensus clustering algorithm is presented to identify a single set of coherent regions valid for all the disturbances considered. Proposed algorithm is implemented to identify coherency in miniWECC 240 bus test system having 20\% of its generation from inverter-based resources.

The proposed consensus clustering identifies the coherency in the presence of IBRs for a wide range of disturbances indicating it's applicability in the high IBR penetrated scenarios. This coherency identification will be instrumental in identifying the required inertial response, fast frequency response, and load repose requirements to address the frequency problems in a more effective way. This coherency identification can be further improved by considering the multi-view aspects as the different variables considered for the coherent identification, variability in the operating conditions of the system, and different methods used for coherency identification.

%The proposed clustering method can prove to be instrumental in identifying a reduced set of representative cases to be examined further for long-term planning studies that focus on grid stability analysis and transient stability studies that are computationally challenging. With the reduced number of cases to be evaluated based on the clustering the generation mix and net-load variation, more detailed grid stability analysis can be performed leading to better planning of the grid to handle interconnection of larger number and higher capacities of variable renewable generation in the system. 

\section*{Acknowledgment}
This work was supported by the Iowa Energy Center (IEC) Grants.

\ifCLASSOPTIONcaptionsoff
  \newpage
\fi

% trigger a \newpage just before the given reference
% number - used to balance the columns on the last page
% adjust value as needed - may need to be readjusted if
% the document is modified later
%\IEEEtriggeratref{8}
% The "triggered" command can be changed if desired:
%\IEEEtriggercmd{\enlargethispage{-5in}}

% ====== REFERENCE SECTION

%\begin{thebibliography}{1}

% IEEEabrv,

\bibliographystyle{IEEEtran}
\bibliography{IEEEabrv,Bibliography}

@ARTICLE{9451565,
  author={Lugnani, Lucas and Paternina, Mario R. Arrieta and Dotta, Daniel and Chow, Joe H. and Liu, Yilu},
  journal={IEEE Transactions on Power Systems}, 
  title={Power System Coherency Detection From Wide-Area Measurements by Typicality-Based Data Analysis}, 
  year={2022},
  volume={37},
  number={1},
  pages={388-401},
  doi={10.1109/TPWRS.2021.3088261}}

@ARTICLE{9141428,
  author={Tyuryukanov, Ilya and Popov, Marjan and van der Meijden, Mart A. M. M. and Terzija, Vladimir},
  journal={IEEE Transactions on Power Systems}, 
  title={Slow Coherency Identification and Power System Dynamic Model Reduction by Using Orthogonal Structure of Electromechanical Eigenvectors}, 
  year={2021},
  volume={36},
  number={2},
  pages={1482-1492},
  doi={10.1109/TPWRS.2020.3009628}}

@ARTICLE{7866837,
  author={Rezaeian, Mohammad Hossein and Esmaeili, Saeid and Fadaeinedjad, Roohollah},
  journal={IEEE Systems Journal}, 
  title={Generator Coherency and Network Partitioning for Dynamic Equivalencing Using Subtractive Clustering Algorithm}, 
  year={2018},
  volume={12},
  number={4},
  pages={3085-3095},
  doi={10.1109/JSYST.2017.2665701}}

@ARTICLE{8708689,
  author={Dabbaghjamanesh, Morteza and Wang, Boyu and Kavousi-Fard, Abdollah and Mehraeen, Shahab and Hatziargyriou, Nikos D. and Trakas, Dimitris N. and Ferdowsi, Farzad},
  journal={IEEE Transactions on Power Delivery}, 
  title={A Novel Two-Stage Multi-Layer Constrained Spectral Clustering Strategy for Intentional Islanding of Power Grids}, 
  year={2020},
  volume={35},
  number={2},
  pages={560-570},
  doi={10.1109/TPWRD.2019.2915342}}

@ARTICLE{9091154,
  author={Huang, Dan and Sun, Huadong and Zhou, Qinyong and Zhao, Shanshan and Zhang, Jian},
  journal={IEEE Access}, 
  title={An WAMS-Based Online Generators Coherency Identification Approach for Controlled Islanding}, 
  year={2020},
  volume={8},
  number={},
  pages={89630-89642},
  doi={10.1109/ACCESS.2020.2993809}}

@techreport{osti_1094827,
  author       = {Dorfler, Florian and Jovanovic, Mihailo R and Chertkov, Michael and Bullo, Francesco},
  title        = {Sparsity-Promoting Optimal Wide-Area Control of Power Networks},
  institution  = {Los Alamos National Laboratory (LANL), Los Alamos, NM (United States)},
  doi          = {10.2172/1094827},
  url          = {https://www.osti.gov/biblio/1094827},
  place        = {United States},
  year         = {2013},
  month        = {07}}

@ARTICLE{8937835,
  author={Ghosh, Sudipta and El Moursi, Mohamed Shawky and El-Saadany, Ehab F. and Hosani, Khalifa Al},
  journal={IEEE Transactions on Power Systems}, 
  title={Online Coherency Based Adaptive Wide Area Damping Controller for Transient Stability Enhancement}, 
  year={2020},
  volume={35},
  number={4},
  pages={3100-3113},
  doi={10.1109/TPWRS.2019.2961004}}

@article{FT_PMU,
author = {A. Vahidnia  and G. Ledwich  and E. Palmer  and A. Ghosh },
title = {Generator coherency and area detection in large power systems},
journal = {IET Generation, Transmission \& Distribution},
volume = {6},
issue = {9},
pages = {874-883},
year = {2012},
doi = {10.1049/iet-gtd.2012.0091},

URL = {https://digital-library.theiet.org/doi/abs/10.1049/iet-gtd.2012.0091},
eprint = {https://digital-library.theiet.org/doi/pdf/10.1049/iet-gtd.2012.0091}
}

@ARTICLE{DMD_PMU,
  author={Barocio, Emilio and Pal, Bikash C. and Thornhill, Nina F. and Messina, Arturo Roman},
  journal={IEEE Transactions on Power Systems}, 
  title={A Dynamic Mode Decomposition Framework for Global Power System Oscillation Analysis}, 
  year={2015},
  volume={30},
  number={6},
  pages={2902-2912},
  doi={10.1109/TPWRS.2014.2368078}}

@INPROCEEDINGS{EnCluster,
  author={Li, Hongmin and Ye, Xiucai and Imakura, Akira and Sakurai, Tetsuya},
  booktitle={2020 IEEE International Conference on Data Mining (ICDM)}, 
  title={Ensemble Learning for Spectral Clustering}, 
  year={2020},
  volume={},
  number={},
  pages={1094-1099},
  doi={10.1109/ICDM50108.2020.00131}}

@ARTICLE{Couple_CI,
  author={Kim, H. and Jang, G. and Song, K.},
  journal={IEEE Transactions on Power Systems}, 
  title={Dynamic reduction of the large-scale power systems using relation factor}, 
  year={2004},
  volume={19},
  number={3},
  pages={1696-1699},
  doi={10.1109/TPWRS.2004.831697}}

@article{EleMechDist_CI,
title = {Electromechanical distance measure for decomposition of power systems},
journal = {International Journal of Electrical Power \& Energy Systems},
volume = {6},
number = {4},
pages = {249-254},
year = {1984},
issn = {0142-0615},
doi = {https://doi.org/10.1016/0142-0615(84)90007-3},
url = {https://www.sciencedirect.com/science/article/pii/0142061584900073},
author = {M.A. Pai and R.P. Adgaonkar}
}

@ARTICLE{SlowCoh_CI,
  author={You, H. and Vittal, V. and Xiaoming Wang},
  journal={IEEE Transactions on Power Systems}, 
  title={Slow coherency-based islanding}, 
  year={2004},
  volume={19},
  number={1},
  pages={483-491},
  doi={10.1109/TPWRS.2003.818729}}

@ARTICLE{RT_CI,
  author={Wei, Jin and Kundur, Deepa and Butler-Purry, Karen L.},
  journal={IEEE Transactions on Smart Grid}, 
  title={A Novel Bio-Inspired Technique for Rapid Real-Time Generator Coherency Identification}, 
  year={2015},
  volume={6},
  number={1},
  pages={178-188},
  doi={10.1109/TSG.2014.2341213}}

@ARTICLE{WECC_FC,
  author={Liu, Shengyuan and Lin, Zhenzhi and Zhao, Yuxuan and Liu, Yilu and Ding, Yi and Zhang, Bo and Yang, Li and Wang, Qin and White, Samantha Emma},
  journal={IEEE Transactions on Power Systems}, 
  title={Robust System Separation Strategy Considering Online Wide-Area Coherency Identification and Uncertainties of Renewable Energy Sources}, 
  year={2020},
  volume={35},
  number={5},
  pages={3574-3587},
  doi={10.1109/TPWRS.2020.2971966}}

\vfill

% Can be used to pull up biographies so that the bottom of the last one
% is flush with the other column.
%\enlargethispage{-5in}

% that's all folks
\end{document}